# Time-domain thermoreflectance (TDTR) measurements of anisotropic thermal conductivity using a variable spot size approach

Puqing Jiang, Xin Qian and Ronggui Yang[a]

*Department of Mechanical Engineering, University of Colorado, Boulder, Colorado 80309, USA*

It is challenging to characterize thermal conductivity of materials with strong anisotropy. In this work, we extend the time-domain thermoreflectance (TDTR) method with a variable spot size approach to simultaneously measure the in-plane ($K_r$) and the through-plane ($K_z$) thermal conductivity of materials with strong anisotropy. We first determine $K_z$ from the measurement using a larger spot size, when the heat flow is mainly one-dimensional along the through-plane direction, and the measured signals are sensitive to only $K_z$. We then extract the in-plane thermal conductivity $K_r$ from a second measurement using the same modulation frequency but with a smaller spot size, when the heat flow becomes three-dimensional, and the signal is sensitive to both $K_r$ and $K_z$. By choosing the same modulation frequency for the two sets of measurements, we can avoid potential artifacts introduced by the frequency-dependent $K_z$, which we have found to be non-negligible, especially for some two-dimensional layered materials like $MoS_2$. After careful evaluation of the sensitivity of a series of hypothetical samples, we provided a guideline on choosing the most appropriate laser spot size and modulation frequency that yield the smallest uncertainty, and established a criterion for the range of thermal conductivities that can be measured reliably using our proposed variable spot size TDTR approach. We have demonstrated this variable spot size TDTR approach on samples with a wide range of in-plane thermal conductivity, including fused silica, rutile titania ($TiO_2$ [001]), zinc oxide (ZnO [0001]), molybdenum disulfide (MoS$_2$), hexagonal boron nitride (*h*-BN), and highly ordered pyrolytic graphite (HOPG).

---

[a] Author to whom correspondence should be addressed. Electronic mail: Ronggui.Yang@Colorado.Edu

## I. INTRODUCTION

Materials with anisotropic thermal conductivity are commonly found in a wide range of condensed matters where the anisotropy is present in the Bravis lattices (ZnO, Ga$_2$O$_3$, etc.), interatomic strengths (layered materials like graphite, chalcogenides, etc.) or nano/micro structures (superlattices, woods, etc.). Understanding anisotropic thermal transport in these materials is not only fundamentally important but also critical to many technological applications including electronics,[1] optoelectronics,[1,2] thermoelectrics,[3-5] thermal insulating[6] and thermal management.[7] However, accurately measuring the thermal conductivities of materials with strong anisotropy remains a great challenge, despite the rapid development in the measuring techniques.[8]

Several techniques could potentially be applied to measure anisotropic thermal conductivities.[8,9] Traditionally, anisotropic thermal conductivities are measured using the steady-heat-flow method, in which many samples are cut with different orientations and then measured.[10,11] This technique, however, requires the samples to be large enough to accommodate at least two thermocouples to measure the temperature gradient, rendering it not suitable to measure the through-plane thermal conductivity of thin films. Over the last three decades, significant progress has been made in using the 3-omega method and the thermoreflectance method, including both the time-domain thermoreflectance (TDTR) and frequency-domain thermoreflectance (FDTR), for measuring thermal conductivity of small samples. In both methods, the samples are heated periodically at the surface, either electrically by a metal line (the 3-omega method) or optically by a laser beam (the thermoreflectance method). Through the variations of the heater size (the metal line width or the laser spot size), both technique can be used to measure anisotropic thermal conductivity, however, with very different sensitivities and complexity in the sample preparation and data reduction. By using different heater line widths or using multiple metal lines (one as a heater and the others as sensors), the 3-omega method has been used extensively to measure anisotropic thermal transport in both bulk and thin film materials.[12,13] However, one significant limitation of the 3-omega method is that it requires not only complicated nano-fabrication of the metal strips but also usually a large and flat sample surface to accommodate them. In comparison, the thermoreflectance method is more flexible, requiring only an optically smooth area of < 100 x 100 $\mu m^2$. Most of the past works used the thermoreflectance method to measure the through-plane thermal conductivity and the interface thermal conductance. Recently a beam-offset TDTR approach developed by Feser *et al.*[14,15] enables independent determination of the in-plane thermal conductivity by using spatially offset pump and probe beams. The beam-offset TDTR approach, however, suffers from large uncertainties, especially when measured with the most commonly used Al transducer, due to the very high sensitivities to both the laser spot sizes and the large thermal conductivity of the transducer film. Rodin and Yee[16] further combined the beam-offset approach with the FDTR to measure the anisotropic thermal conductivity of quartz, sapphire and HOPG, based on the assumption that the through-plane thermal conductivity is independent of the modulation frequency. Such an assumption, however, is not always valid, as the non-equilibrium thermal transport between different heat conduction channels could induce frequency dependence in the apparent through-plane thermal conductivity of some materials, such as SiGe alloy, $MoS_2$ and black phosphorus, as measured by TDTR / FDTR.[17-19]

In this paper, we extend the TDTR with a variable spot size approach to simultaneously measure both the through-plane ($K_z$) and the in-plane ($K_r$) thermal conductivity of bulk materials with anisotropy. We first determine $K_z$ from the measurement using a larger spot size, where the measurement is mostly sensitive to the through-plane thermal transport. We then extract the in-plane thermal conductivity $K_r$ from a second measurement using the same modulation frequency but with a smaller spot size where the measurement is sensitive to both $K_z$ and $K_r$. By choosing the same modulation frequency for the two sets of measurements, we can avoid the error in determining both $K_r$ and $K_z$ introduced by the frequency-dependent $K_z$, which is non-negligible when non-equilibrium thermal transport is pronounced.[17,19] Since we use concentrically aligned pump and probe beams for the TDTR measurements, we confine our work on the samples that are transversely isotropic (with the same in-plane thermal conductivity $K_r$ along different in-plane directions), with a predetermined heat capacity. Some alternative approaches of TDTR, like the beam-offset TDTR,[15] the frequency-dependent TDTR,[20,21] and the dual-frequency TDTR,[22] could be applied for samples with anisotropic $K_r$, unknown heat capacity, and thermally thin films[23], respectively. In Sec. II, we further develop this variable spot size TDTR approach through sensitivity analysis, providing guidelines on choosing the most appropriate laser spot sizes and the modulation frequencies. In Sec. III, experimental data on samples with a wide range of in-plane thermal conductivity are presented to demonstrate the capability of this novel variable spot size TDTR approach for measurement of anisotropic thermal conductivity.

## II. METHODOLOGY

### A. Time-domain thermoreflectance (TDTR) method with the variable spot size approach

Time-domain thermoreflectance (TDTR) method is a robust and powerful technique that can measure thermal properties of a wide variety of materials.[24,25] Our TDTR setup is similar to those in other laboratories.[22,26-28] A schematic diagram of our TDTR system is shown in Fig. 1 (a). In our TDTR setup, we use a mode-locked Ti:sapphire laser that emits a train of 150 fs pulses at 81 MHz repetition rate. The laser beam is corrected to circular shape by a pair of cylindrical lenses before being split into a pump and a probe beam. The pump beam is modulated at a frequency in the range 0.2 – 20 MHz using an electro-optic modulator (EOM). The probe beam is reflected back and forth for three rounds over a 600-mm-long delay stage to achieve up to 12 ns delay time with respect to the pump. The probe beam is expanded before and compressed after the delay stage to make sure the beam size changes by < 2% over the whole delay time range. The pump and the probe beams are then directed into an objective lens and focused concentrically on the sample surface. In our setup, we make the pump and the probe paths parallel but vertically separated by ~4 mm before entering the objective lens. Such spatial separation of the optical paths allows us to use an iris to block the reflected pump beam while allowing the reflected probe beam to pass through into the photodiode detector. We take the ratio between the in-phase ($V_{in}$) and the out-of-phase ($V_{out}$) voltages from the lock-in output, $R = -V_{in} / V_{out}$, as the measured signals, and fit them to a heat transfer model, from which the unknown thermal properties are extracted. The $1/e^2$ radii ($w_0$) of the laser spots at the sample surface are varied in the range 4 – 40 μm, achieved by choosing objective lenses with different magnifications. In our setup, the spot size of the pump is slightly larger (~10%) than that of the probe. However, what really matters in TDTR is the root-mean-square (RMS) average of the pump and the probe spot sizes; therefore, some slight difference between them should not matter. The laser spot sizes $w_0$ mentioned in this work, if not specifically indicated, all represent the RMS average of the pump and the probe.

TDTR can be used to measure multiple thermal properties under different heat transport regimes, achieved by changing the laser spot size and the modulation frequency.[20,25] There are two important length scales that determine the heat transport regime, i.e., the laser spot size $w_0$ that affects the lateral heat spreading, and the through-plane thermal diffusion length under periodic heating, also known as the thermal penetration depth $d_{p,z}$, defined as $d_{p,z} = \sqrt{K_z/\pi f C}$, with $K_z$ the through-plane thermal conductivity, $f$ the modulation frequency and $C$ the volumetric heat capacity. When TDTR experiment is conducted using a laser spot size much larger than the through-plane (z-) thermal penetration depth, the temperature gradient is mainly one-dimensional along the through-plane direction, see Fig. 1 (b). In such a configuration, the detected surface temperature change is predominantly affected by the through-plane thermal conductivity. On the other hand, when the measurement is conducted using a tightly focused laser spot whose size is comparable to the thermal penetration depth, the temperature gradient becomes three-dimensional, as shown in Fig. 1 (c). Under such a heat transport regime, the measured surface temperature change depends on both $K_r$ and $K_z$. We can thus determine both $K_r$ and $K_z$ by conducting two sets of measurements using different laser spot sizes. Note that the through-plane thermal penetration depth can also be controlled by choosing different modulation frequencies, with $d_{p,z} \sim f^{-1/2}$. Thus, the conventional practice for anisotropic measurements using TDTR is to determine $K_z$ first by using a large spot size at a high modulation frequency, and then to measure $K_r$ using a small spot size at a low modulation frequency, by using the measured $K_z$ as an input for the heat transfer model.[25] However, we should be cautious because this variable spot size approach is in fact based on the assumption that $K_z$ is the same for the two sets of measurements, while some previous experiments[18,19,29,30] and theoretical analyses[17,31] suggest that $K_z$ of some materials could depend on the modulation frequency.

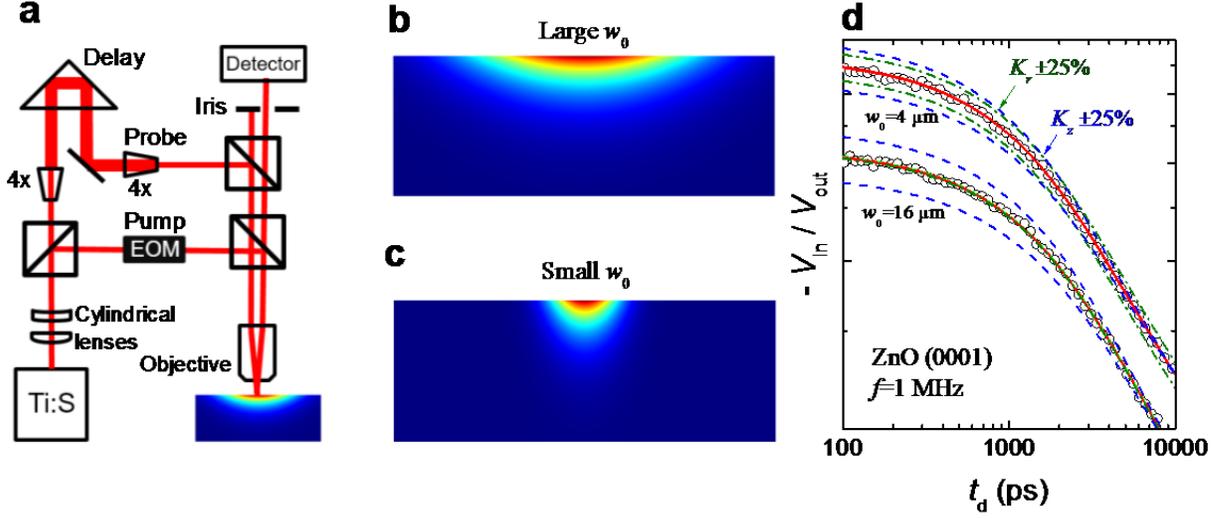

FIG. 1. (a) Schematic of TDTR setup. (b and c) Different heat transport regimes achieved by using a large laser spot size (which is mainly one-dimensional), and by using a tightly focused spot size (which is three-dimensional). (d) Representative data fitting using the variable spot size TDTR approach to measure anisotropic thermal conductivity of ZnO [0001].

To avoid the possible error in determining both $K_r$ and $K_z$ resulted from the frequency-dependent $K_z$, the same modulation frequency should be chosen to perform the variable spot size measurements. Figure 1 (d) shows representative experimental signals $R = -V_{in} / V_{out}$ and their fitting curves for ZnO [0001] as a function of delay time between the pump and the probe beams. The data in this figure were taken at room temperature using two different laser spot sizes ($1/e^2$ radius $w_0$ = 4 μm and $w_0$ = 16 μm) under the same modulation frequency of 1 MHz. The measurement using the large spot size ($w_0$ = 16 μm) is only sensitive to the through-plane thermal conductivity, which is extracted as $K_z$ = 55 ± 6 W m$^{-1}$ K$^{-1}$ from the fitting of the heat transfer model to the experimental signal. With the through-plane thermal conductivity determined, we then extract the in-plane thermal conductivity as $K_r$ = 42 ± 10 W m$^{-1}$ K$^{-1}$ from the second measurement using a smaller spot size ($w_0$ = 4 μm) at the same modulation frequency. See Section II B for details of the uncertainty analysis.

Although the physical picture for anisotropic thermal conductivity measurements using the variable spot size TDTR approach is straightforward, some essential questions remain unanswered. For example, how should we choose the most appropriate modulation frequency and laser spot sizes to yield the smallest measurement uncertainty? What is the thermal conductivity range that we can measure with an acceptable error bar using this variable spot size approach? To answer these questions, we conduct sensitivity analysis on a series of hypothetical samples, from which we can draw some conclusions and provide the guidelines on the variable spot size approach.

Whether TDTR can be used to measure an unknown thermal property with confidence or not depends on the sensitivity of the measured signal to that unknown property. The sensitivity coefficient is the dimensionless partial derivative of the measured signal (the ratio $R = -V_{in}/V_{out}$) with respect to the unknown parameter $\alpha$:[32]

$$S_\alpha = \frac{\alpha}{R}\frac{\partial R}{\partial \alpha} \tag{1}$$

The sensitivity coefficient $S_\alpha$ and its sign have distinctive meanings. For example, a sensitivity of $S_\alpha = 0.5$ means that a 10% increase in the parameter $\alpha$ will result in 5% increase in the signal $R$. The sensitivity ratio $S_\alpha/S_\beta$, on the other hand, describes the error propagation between the two parameters $\alpha$ and $\beta$. For example, a sensitivity ratio of $S_\alpha/S_\beta = 2$ means that a 10% uncertainty in $\alpha$ and a 20% uncertainty in $\beta$ would cause the same amount of change in $R$. In other words, the measurement is more sensitive to parameter $\alpha$ than $\beta$.

Based on the discussion above, we use the sensitivity ratio $S_{Kz}/S_{Kr}$ as the criterion to determine the workable modulation frequency range for anisotropic thermal conductivity measurement using the variable spot size approach. Figure 2(a) shows an example of the sensitivity ratio $S_{Kz}/S_{Kr}$ of a hypothetical sample (the properties are taken as $K_r$ = 100 W m$^{-1}$ K$^{-1}$, $K_z$ = 5 W m$^{-1}$ K$^{-1}$, $C$ = 2 MJ m$^{-3}$ K$^{-1}$ and $G$ = 50 MW m$^{-2}$ K$^{-1}$, which are the typical values of anisotropic materials), plotted as a function of modulation frequency using two different spot sizes ($w_0$=5 µm and 20 µm). To determine the through-plane thermal conductivity $K_z$ independently without any prior knowledge of the in-plane thermal conductivity $K_r$, we need a large sensitivity ratio $S_{Kz}/S_{Kr}$ > 10 so that even 20% error in $K_r$ would contribute < 2% error in $K_z$. Similarly, we require the sensitivity ratio $S_{Kz}/S_{Kr}$ < 2 so that a ~10% error in $K_z$ would result in < 20% uncertainty in $K_r$. Considering the extra uncertainty contributed from other input parameters ($h_{Al}$, $C_{Al}$, $w_0$, etc.), we should be able to determine $K_r$ with a < 30% uncertainty. We thus determine the lower frequency limit $f_L$ using the criterion $S_{Kz}/S_{Kr}$ > 10 from the curve using the larger spot size ($w_0$ = 20 µm), and the upper frequency limit $f_U$ using the criterion $S_{Kz}/S_{Kr}$ < 2 from the curve using the smaller spot size ($w_0$ = 5 µm). To avoid the potential artifacts from the frequency-dependent $K_z$, we should choose a single modulation frequency in the range $f_L < f < f_U$ to determine both $K_z$ and $K_r$. Measurements outside this modulation frequency range could not determine $K_z$ independently, or determine $K_r$ with an acceptable uncertainty (>30%).

However, the thermal properties of real materials spread over a few orders of magnitude. A more generic guidance is thus necessary on choosing the appropriate modulation frequency based on the thermal properties of the material. We therefore studied a series of hypothetical samples over a wide range of thermal properties: $K_z$ = 0.1 – 1000 W m$^{-1}$ K$^{-1}$, $K_r$ = 1 – 1000 W m$^{-1}$ K$^{-1}$, $K_r$ / $K_z$ = 0.01 – 10000, $C$ = 0.3 – 3 MJ m$^{-3}$ K$^{-1}$, and interfacial thermal conductance $G$ = 30 – 1000 MW m$^{-2}$ K$^{-1}$. For each combination of the thermal properties ($K_r$, $K_z$, $C$ and $G$), we performed the similar sensitivity analysis as shown in Fig. 2 (a), from which we picked up the frequency limits $f_L$ and $f_U$, and compiled them in Fig. 2 (b). We found that the frequency limits $f_L$ and $f_U$ determined from the sensitivity analysis predominantly depend on the in-plane thermal diffusivity $K_r / C$, despite the wide range of thermal properties, as shown in Fig. 2 (b). From such numerical experiments, we extract an empirical correlation between the workable modulation frequency $f$ and the in-plane thermal diffusivity $K_r / C$ as

$$f = a(K_r/C)^{0.7} \tag{2}$$

with $f$ having a unit of MHz and $K_r / C$ a unit of cm$^2$ s$^{-1}$, and the constant $a$ in the range 2.2 – 4.4.

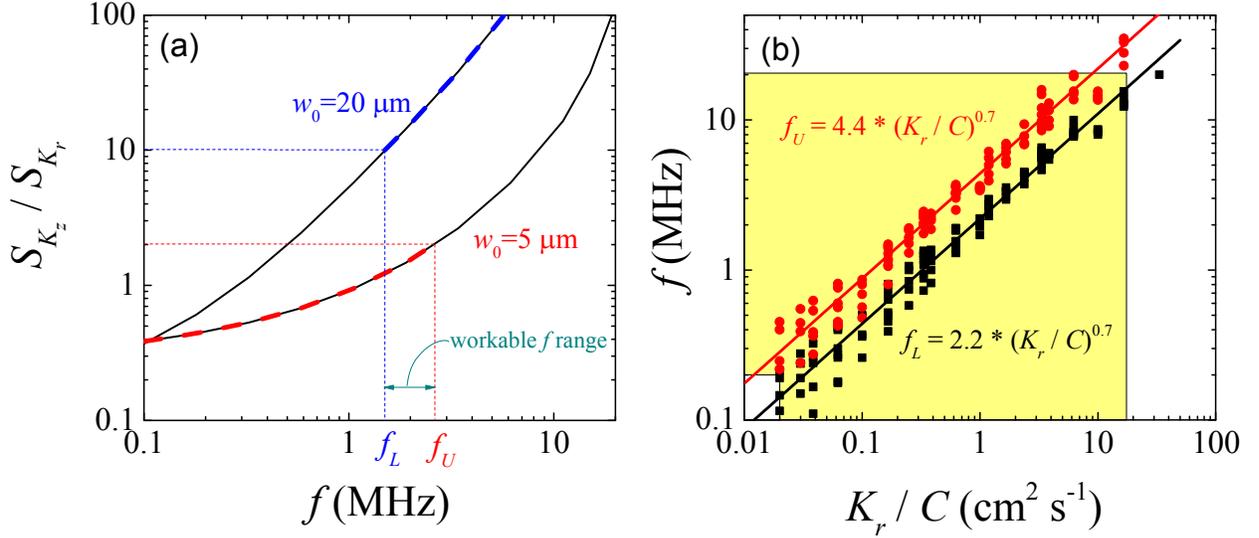

FIG. 2. (a) The workable frequency range for the variable spot size approach determined from the frequency-dependent sensitivity ratio $S_{Kz}/S_{K_r}$, when measured using two different laser spot sizes ($w_0 = 5$ μm and $w_0 = 20$ μm). (b) The correlation between the workable frequency range for the variable spot size approach and the in-plane thermal diffusivity $K_r / C$, obtained from the calculations on a series of hypothetical samples over a wide range of thermal properties.

Figure 2 (b) also helps us to determine that the measurable $K_r / C$ using the variable spot size

TDTR approach. Considering that TDTR works the best at frequencies 0.2 – 20 MHz, the measurable $K_r / C$ should be in the range 0.02 – 20 cm$^2$ s$^{-1}$. TDTR measurements at $f < 0.2$ MHz are usually problematic due to the poor signal-to-noise ratio caused by the $1/f$ noise and the large uncertainty in determining the phase, while measurements at $f > 20$ MHz are also challenging due to the weak out-of-phase signals and the high level of the radio-frequency noise picked up by the detector and the signal cables.[33] We note that the empirical correlations we provided in Fig. 2 (b) is only meant to provide a rough guideline on the workable modulation frequency range. The most appropriate modulation frequency for the variable spot size approach should still be determined on a case-by-case basis. For example, the lower limit $f_L$ in Fig. 2 (b) was determined with a spot size $w_0 = 20$ μm; this limit can be even lower when using a larger laser spot.

Using the similar sensitivity analysis, we can also determine the most appropriate laser spot size, as shown in Fig. 3. Figure 3 (a) shows the contour of the sensitivity ratio $S_{Kz}/S_{Kr}$ as a function of the laser spot size $w_0$ and the in-plane thermal penetration depth $d_{p,r}$, which has the same definition as $d_{p,z}$, except to replace $K_z$ with $K_r$. Based on the criterion that we can independently measure $K_z$ when $S_{Kz}/S_{Kr} > 10$, we need a laser spot size $w_0 > 5d_{p,r}$. On the other hand, we need a small spot size $w_0 < 2d_{p,r}$, when we have the sensitivity ratio $S_{Kz}/S_{Kr} < 2$, to measure $K_r$ with an acceptable uncertainty. To further illustrate the effect of the laser spot size on the measurement uncertainty of $K_r$, we calculated the uncertainty in $K_r$ of ZnO [0001] when measured using different spot sizes $w_0 = 1 - 10$ μm at a fixed modulation frequency of 1 MHz, with $K_z$ pre-determined with ~11% uncertainty from a separate measurement using a large spot size $w_0 = 16$ μm. Details on the methods of uncertainty analysis can be found in Section II B. We found that we need a small spot size of $w_0 < 5$ μm (equivalently $w_0 < 2d_{p,r}$) to have a <

30% uncertainty for $K_r$, see Fig. 3 (b). We also found that pushing the laser spot size to the smaller limit below 3 μm would not help improve the uncertainty of $K_r$, because the benefit from the increased sensitivity to $K_r$ at smaller spot sizes is counteracted by the similarly increased sensitivity to $w_0$.

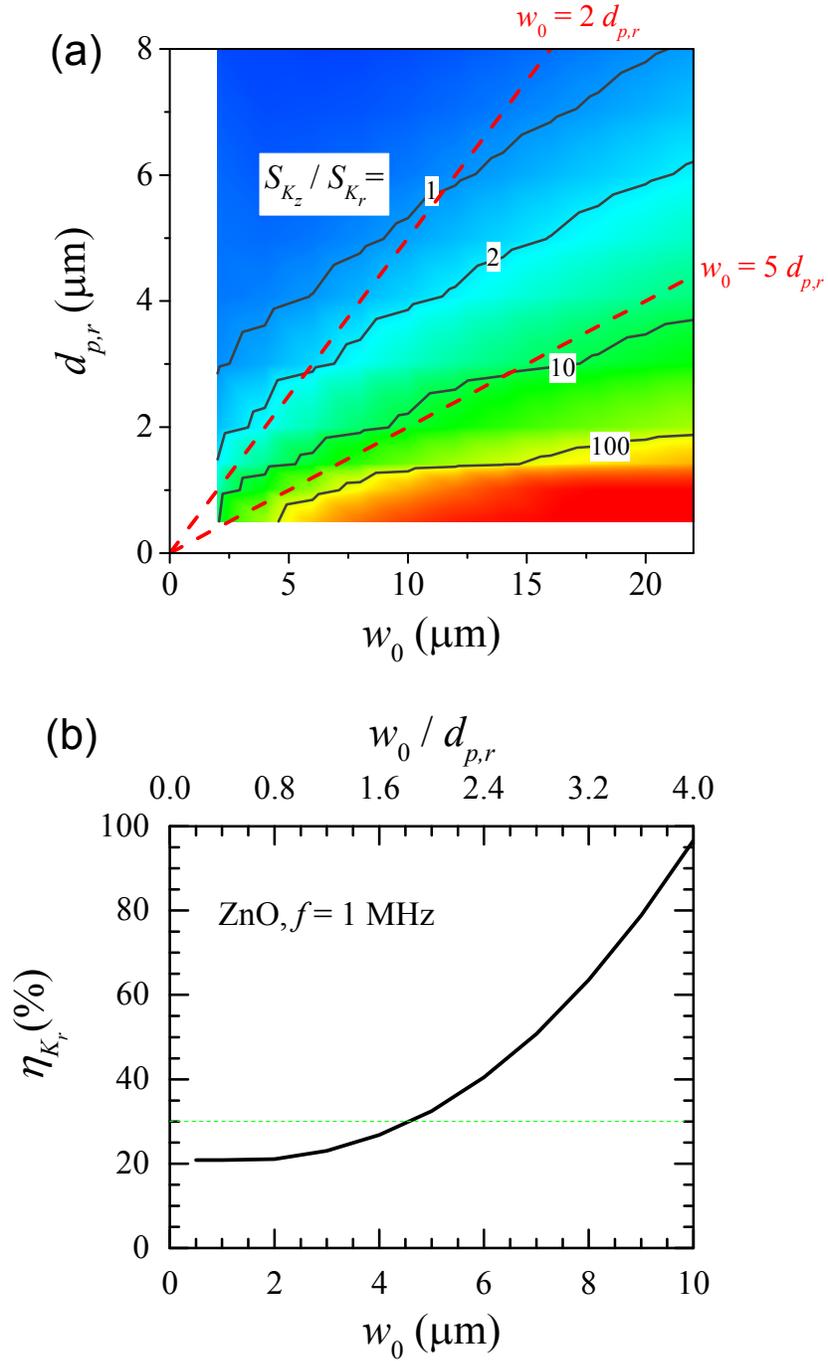

FIG. 3. (a) The contour of the sensitivity ratio of $K_z$ to $K_r$ as a function laser spot size $w_0$ and in-plane thermal penetration depth $d_{p,r}$, from which we can determine the most appropriate laser spot sizes for the variable spot size approach. (b) Uncertainty of $K_r$ of ZnO measured as a function of laser spot size.

## B. Uncertainty analysis

Usually the standard deviation method is used for uncertainty analysis in TDTR experiments,[21,22,34] which assumes independent uncertainties for the different input parameters. However, this standard deviation method cannot be used in our case since the uncertainty of $K_r$ depending on that of $K_z$, while the uncertainties of both $K_r$ and $K_z$ depend on the uncertainties of the other input parameters. Besides, since we simultaneously determine both the interface conductance $G$ and the substrate thermal conductivity $K_z$ or $K_r$ from TDTR measurements,[23] the effects of the uncertainty from $G$ on the uncertainty of $K_z$ and $K_r$ and vice versa are not clear if the uncertainties are estimated using the standard deviation method.

In recent years, the Monte Carlo technique has emerged as a more conservative yet straightforward approach to estimate experimental uncertainty.[16,35,36] The great advantage of the Monte Carlo method over the analytical standard derivative method in uncertainty estimation is that it requires no prior knowledge of how the uncertainties of different input parameters may interact; thus, the Monte Carlo method is more suitable for our case. To implement the Monte Carlo method for uncertainty analysis, we assume each input parameter to have a normal distribution about its nominal value, with the uncertainty (2σ, 95% confidence level) determined from separate measurements. We randomly select a set of values for the input parameters from their distributions, and fit the experimental data measured using the large spot size for $G$ and $K_z$.

The uncertainty from $K_r$ should not matter for this case since the measurement using the large spot size is not sensitive to $K_r$. We then use the same set of the input parameters and the fitted $G$ and $K_z$ values to fit the second experimental data measured using the small spot size for $K_r$. This process follows exactly the actual data processing procedure in our variable spot size measurements. We fit the interface conductance $G$ from the measurement using the large spot size because in such a configuration the heat flow is mainly one-dimensional across the interface (in the through-plane direction), the measured data is more sensitive to $G$, and consequently we can determine $G$ with a smaller uncertainty. We repeat this process 5000 times to generate a distribution of the possible outcomes for $G$, $K_z$ and $K_r$, from which we can estimate the uncertainty of $G$, $K_z$ and $K_r$ based on the 95% confidence level. Among the input parameters, we assume an uncertainty of 10% for $K_{Al}$, 3% for $C_{Al}$ and $C_{sub}$, 4% for $h_{Al}$, and 5% for $w_0$.

An example of the Monte Carlo histograms for the uncertainty estimation of $K_z$, $K_r$ and $G$ is shown in Fig. 4 for ZnO [0001], measured at 1 MHz using two different spot sizes ($w_0$ = 16 μm and $w_0$ = 4 μm). From the histograms, we estimate the uncertainty to be ~10% and ~16% for $K_z$ and $K_r$ of ZnO, respectively, and ~4% for the Al/ZnO interface thermal conductance $G$. The estimated uncertainties of all the samples measured are tabulated in Table I.

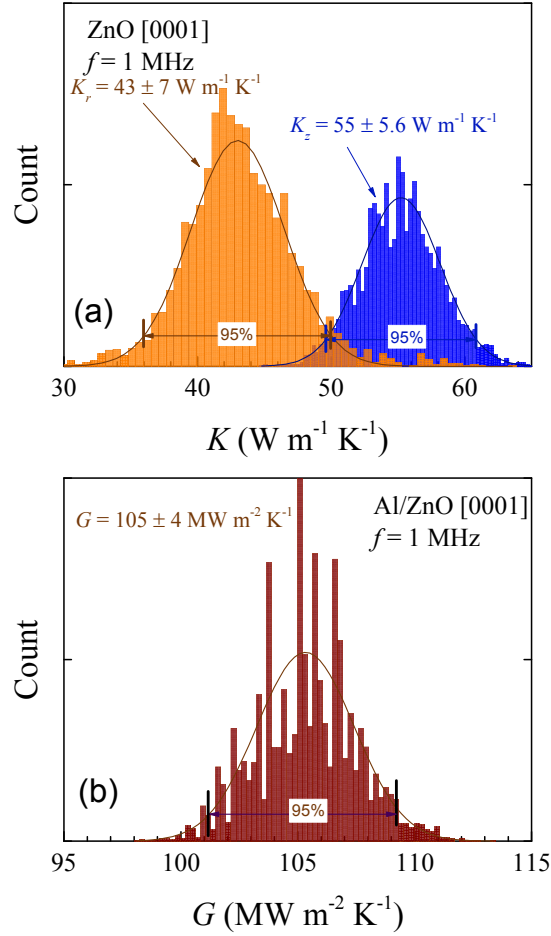

FIG. 4. Monte Carlo histograms for uncertainty estimations of $K_z$, $K_r$ and $G$ for ZnO [0001].

## C. Sample preparation

We choose six samples that cover a wide range of in-plane thermal diffusivity (conductivity) and degrees of anisotropy to demonstrate the capability of the variable spot size TDTR approach for simultaneous measurements of $K_r$ and $K_z$. These samples are fused silica, rutile titania (TiO$_2$ [001]), zinc oxide (ZnO [0001]), molybdenum disulfide (MoS$_2$), hexagonal boron nitride ($h$-BN), and highly ordered pyrolytic graphite (HOPG), with their nominal thermal conductivity and volumetric heat capacity values listed in Table I.

Of these samples, the silica, TiO$_2$ and ZnO wafers were purchased from MTI, while the MoS$_2$ crystals (naturally mined), *h*-BN single crystals (grade A) and HOPG (grade SPI-1) were purchased from SPI Supplies®. Among these samples, TiO$_2$ and ZnO have a higher thermal conductivity along the c-axis than that in other directions, while MoS$_2$, *h*-BN and HOPG are layered materials that have a much larger thermal conductivity along the in-plane direction than the through-plane direction. We chose the TiO$_2$ and ZnO crystals to have their c-axis along the z-direction, so that their lateral (in-plane) thermal conductivities are isotropic. The *h*-BN crystals come as 1-mm-sized flakes with a thickness > 10 μm, since the growth of high-quality large-sized *h*-BN crystals is very difficult.[37] We believe that TDTR is advantageous over other techniques, like the steady-heat-flow method, laser flash method and the 3ω method, to measure the thermal conductivity of such small-sized samples, because TDTR only requires a small area of 100 x 100 μm$^2$ for the measurement. To ease the TDTR measurements, we glue the small *h*-BN crystals on a large Si wafer using silver paste. Since the thermal penetration depth in *h*-BN using the appropriate modulation frequency during the TDTR measurements is only < 1 μm, much smaller than the thickness of the *h*-BN crystals (> 10 μm), we can consider the *h*-BN crystals as semi-infinite solids, with the effect of the silver paste and the Si substrate safely ignored.

To prepare the samples for TDTR measurements, we deposit a layer of 100 nm Al film on the samples as a transducer. The silica, TiO$_2$ and ZnO wafers were cleaned from any organic residue using isopropyl alcohol and ethanol, while the first few layers of MoS$_2$, *h*-BN and HOPG samples were exfoliated away using a Scotch tape before the deposition of Al transducers.

**III. RESULTS AND DISCUSSIONS**

As a demonstration of the variable spot size TDTR approach, we measured $K_r$, $K_z$ and $G$ of fused silica, $TiO_2$ [001], ZnO [0001], $MoS_2$, $h$-BN, and HOPG using Al transducer, with their heat capacity obtained from literature. We check first the possible frequency dependence in $K_z$ and $G$ of the samples, measured using a large spot size $w_0 = 20$ μm, as shown in Fig. 5. We find that except for $MoS_2$, which shows a ~30% decrease in $K_z$ and a ~30% increase in $G$ as the frequency increases from 0.2 MHz to 10 MHz, none of the other samples show any frequency dependence in either $K_z$ or $G$. Similar frequency dependence in both $K_z$ and $G$ of $MoS_2$ has also been previously reported on SiGe alloy and was analyzed using a two-channel model.[17] The reason for the frequency dependence in both $K_z$ and $G$ for $MoS_2$, as we have recently analyzed in another paper,[19] is that the non-equilibrium thermal resistance between different heat conduction channels in $MoS_2$ near the interface manifests in different manners at different modulation frequencies. When measured using a very low modulation frequency with a very long thermal penetration depth, the non-equilibrium only happens near the interface, resulting in a lower apparent thermal conductance $G$. At higher modulation frequencies with shorter thermal penetration depths, the non-equilibrium takes into effect throughout the whole thermally excited region, resulting in lower apparent values for both $G$ and $K_z$. Since $MoS_2$ shows frequency dependence in both $K_z$ and $G$, we obtain the intrinsic values by analyzing the experimental data using a two-channel model, see Refs[17,19] for more details. We note that although $h$-BN and HOPG have the same van-der-Waals structures as $MoS_2$, they do not show any frequency dependence in either $K_z$ or $G$. The difference could possibly be attributed to a big phonon bandgap that only exists in $MoS_2$ but does not present in either $h$-BN or HOPG. The big phonon bandgap results in weak coupling and consequently large non-equilibrium thermal resistance between the different heat conduction channels in $MoS_2$. The frequency dependence in $K_z$ and $G$ of MoS$_2$ supports our argument that we should choose the same modulation frequency to avoid possible errors associated with the frequency-dependent $K_z$, when using the variable spot size approach to simultaneously measure $K_r$ and $K_z$.

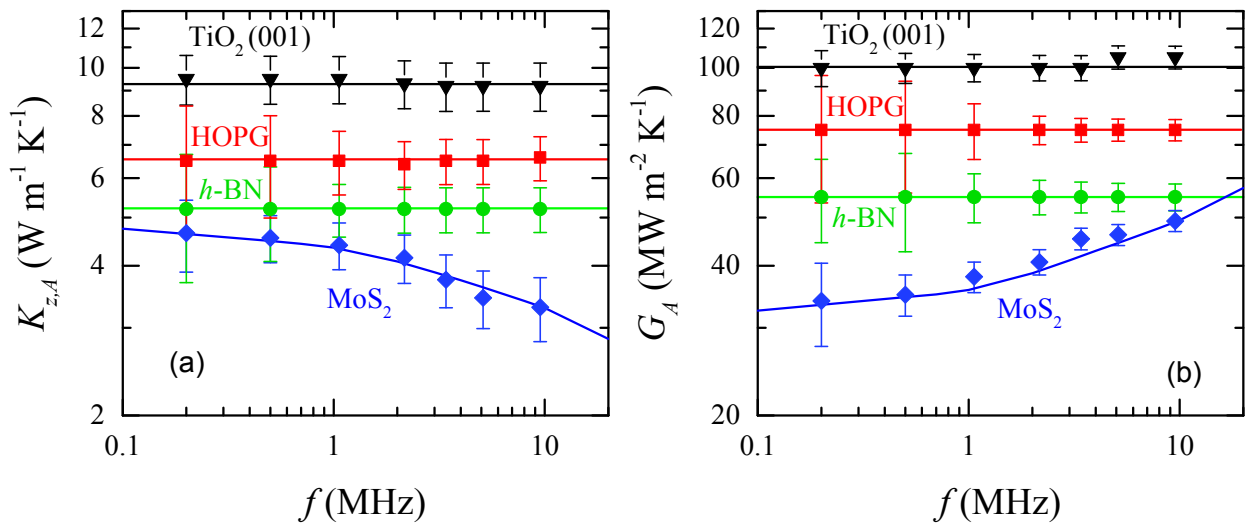

FIG. 5. *Apparent through-plane thermal conductivity $K_z$ and interface conductance G of TiO$_2$ [001], HOPG, h-BN and MoS$_2$, measured as a function of modulation frequency, using a large laser spot size $w_0$ = 20 μm. Symbols represent the TDTR measurements and the solid lines indicate predicted results using a two-channel model.*

We summarize our measured in-plane and through-plane thermal conductivity of the six samples in Fig. 6, compared with the literature values. All the data are also tabulated in Table I, along with the known heat capacities obtained from literature. These samples cover a wide range of the in-plane thermal diffusivity 0.0084 – 11.1 cm$^2$ s$^{-1}$. Both our measured $K_r$ and $K_z$ of these samples compare very well with the literature values, which validates our variable spot size TDTR approach. Overall, our measured $K_r$ have uncertainties < 20%, except for TiO$_2$, which has an uncertainty of ~30%, and the fused silica, which has an uncertainty of 160% in $K_r$.

Considering that silica has a low in-plane thermal diffusivity of 0.0084 cm$^2$ s$^{-1}$, the exceptionally large uncertainty for $K_r$ of silica is consistent with our previous analytical results that we could only measure $K_r$ of the samples with in-plane thermal diffusivity $K_r / C$ in the range 0.02 – 20 cm$^2$ s$^{-1}$ using our variable spot size TDTR approach. Our measured $K_z$ of these samples generally have an uncertainty of ~11%. Note that although the through-plane thermal conductivity of these samples fall in the range 1 – 100 W m$^{-1}$ K$^{-1}$, TDTR has been previously applied to measure the through-plane thermal conductivity across a much wider range from very high thermal conductivity of diamond (~2000 W m$^{-1}$ K$^{-1}$) to ultralow thermal conductivity of disordered layered crystals and fullerene derivatives (~0.03 W m$^{-1}$ K$^{-1}$).[23]

Our measured $G$ of these samples generally have an uncertainty of ~7%, as shown in Table I, except for the Al/silica interface conductance, which has an uncertainty of ~17%. The larger uncertainty of $G$ for silica is easy to understand, as the low thermal conductivity of silica hinders the heat flux from penetrating across the interface into the substrate, thus the measured signal is less sensitive to the interface thermal conductance.

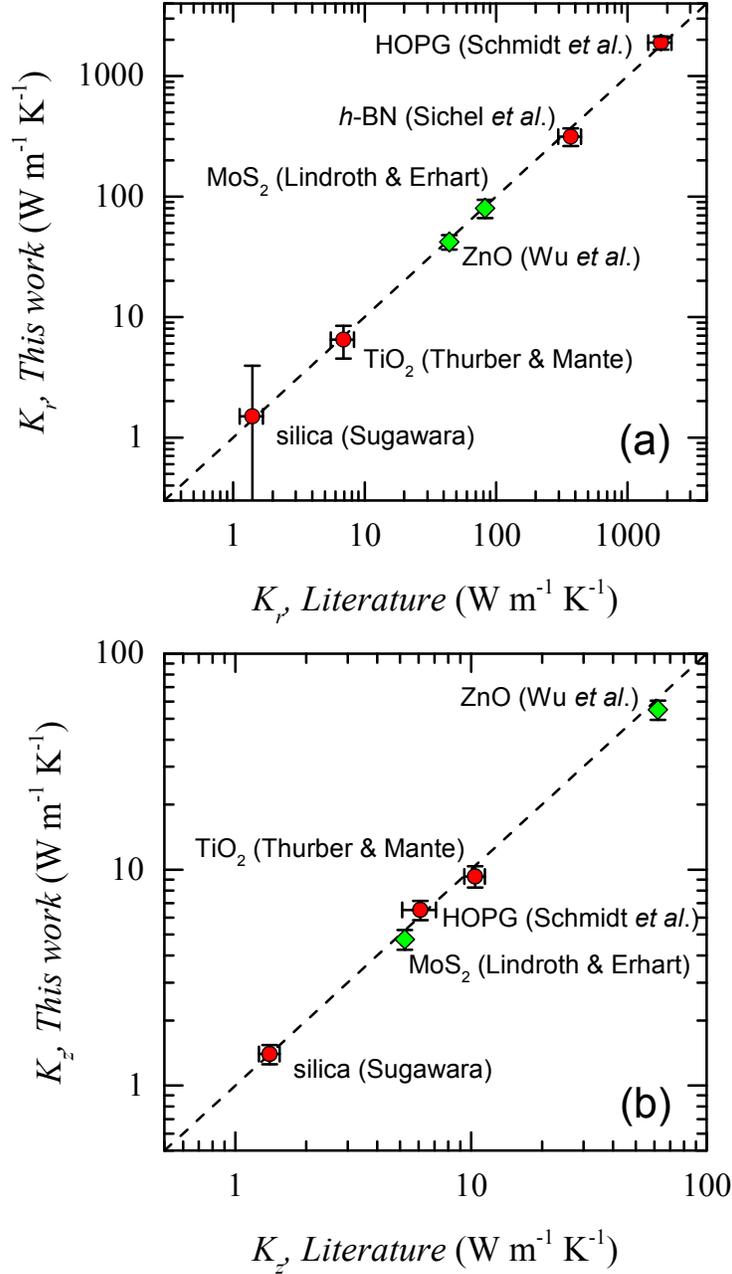

FIG. 6. Comparison of our measured values with the literature for $K_r$ and $K_z$ of silica, $TiO_2$ [001], ZnO [0001], $MoS_2$, h-BN and HOPG. Of the literature values, those for $MoS_2$ and ZnO (diamond symbols) were from first-principles calculations, while the others (circle symbols) were from experiments. Since there is a lack of the literature value for $K_z$ of h-BN, our measured $K_z$ of h-BN is not shown in the plot. The detailed references of the literature values can be found in Table I.

## IV. CONCLUSION

A variable spot size TDTR approach has been developed to simultaneously measure the in-plane ($K_r$) and the through-plane ($K_z$) thermal conductivity of materials with strong anisotropy. In the variable spot size TDTR approach, we first determine $K_z$ from the measurement using a large spot size, when the measured signals are sensitive to only $K_z$; we then extract $K_r$ from a second measurement using the same modulation frequency but with a smaller spot size, when the signal is sensitive to both $K_r$ and $K_z$. By choosing the same modulation frequency for the two sets of measurements, we can avoid potential errors associated with the frequency-dependent $K_z$. We also provided guidelines on choosing the most appropriate laser spot size and modulation frequency that yield the smallest uncertainty, and established a criterion on the range of in-plane thermal conductivities that can be reliably measured using our variable spot size TDTR approach. This variable spot size TDTR approach is demonstrated on samples with a wide range of in-plane thermal conductivity 1 – 2000 W m$^{-1}$ K$^{-1}$, with the measurement uncertainty in $K_r$ generally < 30% when $K_r$ > 6.5 W m$^{-1}$ K$^{-1}$.

**ACKNOWLEDGMENT:**

We acknowledge the preliminary work conducted by Jun Liu and Wei Wang on developing the variable spot size TDTR approach in 2012-2015. This work is supported by the NSF (Award No. 1511195) and DOD DARPA (Grant No. FA8650-15-1-7524).

TABLE I. The literature and measured values for in-plane thermal conductivity ($K_r$), through-plane thermal conductivity ($K_z$), and volumetric heat capacity ($C$), and the measured values for the interface thermal conductance ($G$) between Al and the samples.

| Sample | Literature | | | | Current | | |
| --- | --- | --- | --- | --- | --- | --- | --- |
| | $C$ (J cm$^{-3}$ K$^{-1}$) | $K_r$ (W m$^{-1}$ K$^{-1}$) | $K_z$ (W m$^{-1}$ K$^{-1}$) | $K_r / C$ (cm$^2$ s$^{-1}$) | $K_r$ (W m$^{-1}$ K$^{-1}$) | $K_z$ (W m$^{-1}$ K$^{-1}$) | $G$ (MW m$^{-2}$ K$^{-1}$) |
| Fused silica | 1.65 ± 0.05[38] | 1.39 ± 0.14[39] | 1.39 ± 0.14[39] | 0.0084 | 1.5 ± 2.44 | 1.4 ± 0.15 | 170 ± 30 |
| TiO$_2$ [001] | 2.91 ± 0.09[40] | 7.0 ± 0.5[10] | 10.4 ± 0.7[10] | 0.024 | 6.5 ± 1.98 | 9.3 ± 1.05 | 100 ± 6 |
| ZnO [0001] | 2.81 ± 0.09[41] | 44[42] | 62[42] | 0.16 | 42 ± 5.8 | 55 ± 5.6 | 105 ± 6 |
| MoS$_2$ | 1.91 ± 0.05[43] | 82.3[44] | 5.23[44] | 0.43 | 80 ± 14 | 4.75 ± 0.5 | 50 ± 3 |
| h-BN | 1.77 ± 0.05[45] | 370[46] | -- | 2.03 | 315 ± 52 | 5.2 ± 0.6 | 55 ± 4 |
| HOPG | 1.62 ± 0.05[47] | 1800 ± 200[11,48] | 6.1 ± 1.2[11] | 11.1 | 1900 ± 240 | 6.5 ± 0.7 | 75 ± 4 |